\documentclass[preprint]{vgtc}               % preprint
\ifpdf%                                % if we use pdflatex
  \pdfoutput=1\relax                   % create PDFs from pdfLaTeX
  \usepackage{graphicx}                % allow us to embed graphics files
  \DeclareGraphicsExtensions{.pdf,.png,.jpg,.jpeg} % for pdflatex we expect .pdf, .png, or .jpg files
  \usepackage[noadjust]{cite}
\else%                                 % else we use pure latex
  \usepackage{graphicx}                % allow us to embed graphics files
  \DeclareGraphicsExtensions{.eps}     % for pure latex we expect eps files
\fi%

\graphicspath{{figures/}{pictures/}{images/}{./}} % where to search for the images

\usepackage{microtype}                 % use micro-typography (slightly more compact, better to read)
\PassOptionsToPackage{warn}{textcomp}  % to address font issues with \textrightarrow
\usepackage{textcomp}                  % use better special symbols
\usepackage{mathptmx}                  % use matching math font
\usepackage{times}                     % we use Times as the main font
\usepackage{cite}                      % needed to automatically sort the references
\usepackage{tabu}                      % only used for the table example
\usepackage{booktabs}                  % only used for the table example
\onlineid{1184}

\vgtccategory{Research}

\vgtcinsertpkg

\title{Jurassic Mark: Inattentional Blindness for a Datasaurus Reveals that Visualizations are Explored, not Seen}

\author{Tal Boger\thanks{e-mail: tal.boger@yale.edu}\\ %
        \scriptsize Yale University %
\and Steven B. Most\thanks{e-mail: s.most@unsw.edu.au}\\ %
     \scriptsize UNSW Sydney %
\and Steven L. Franconeri\thanks{e-mail: franconeri@northwestern.edu}\\ %
     \scriptsize Northwestern University}

\abstract{Graphs effectively communicate data because they capitalize on the visual system’s ability to rapidly extract patterns. Yet, this pattern extraction does not occur in a single glance. Instead, research on visual attention suggests that the visual system iteratively applies a sequence of filtering operations on an image, extracting patterns from subsets of visual information over time, while selectively inhibiting other information at each of these moments. To demonstrate that this powerful series of filtering operations also occurs during the perception of visualized data, we designed a task where participants made judgments from one class of marks on a scatterplot, presumably incentivizing them to relatively ignore other classes of marks. Participants consistently missed a conspicuous dinosaur in the ignored collection of marks (93\% for a 1s presentation, and 61\% for 2.5s), but not in a control condition where the incentive to ignore that collection was removed (25\% for a 1s presentation, and 11\% for 2.5s), revealing that data visualizations are not ``seen" in a single glance, and instead require an active process of exploration.} % end of abstract

\begin{document}

%% The ``\maketitle'' command must be the first command after the
%% ``\begin{document}'' command. It prepares and prints the title block.

%% the only exception to this rule is the \firstsection command
\firstsection{Introduction}

\maketitle

When data are represented as effectively designed visualizations, the human visual system can rapidly extract statistics, trends, and other patterns \cite{szafir2016four, healey2011attention, franconeripspi}. While this process can feel instantaneous, a large body of literature from perceptual psychology suggests that some visual tasks, such as finding particular objects in crowded scenes, making comparisons to find differences, or extracting relations among objects, require an iterative set of filtering operations that progressively isolate subsets of visual information \cite{franconeri2013nature}. Most of these tasks are also key to exploring visualized data – inspecting a visualization is not always as instantaneous as recognizing an object, and instead in many cases this process must unfold slowly over time \cite{franconeriCurr}. Depending on how that process unfolds, the same underlying data can be perceived differently, leading designers to strive to create visualizations that help viewers see the ``right" pattern in a dataset \cite{shah2011bar, ajani2021declutter, zacks2020designing, xiongVisual}.

This filtering process can unfold in several ways when looking at a visualization. For example, one may filter for a certain color or shape (filtering in a feature space), bars on a certain side of a plot (filtering in space), or a single line that crosses with another line of the same color (filtering for a given object) \cite{franconeri2013nature, brooks2015traditional}. The need for these progressive filtering steps slows down the otherwise rapid process of recognizing statistics, trends, and other patterns within an image. This filter can be so powerful that it can lead people to surprising failures to process information that is left out by that filter. Here, we illustrate the power of this visual filter by demonstrating a surprising example of such ``inattentional blindness."

Inattentional blindness is a robust phenomenon in which people fail to see salient objects that are blocked by their current visual filter. In most cases, inattentional blindness is demonstrated by showing that people can miss surprisingly salient real-world objects within otherwise simple scenes, such as an additional person leisurely walking through, or even a gorilla \cite{neisser1975selective, simons1999gorillas}. Other work has also found inattentional blindness where people fail to notice (or fail to have their performance be influenced by) salient shapes or groupings that are created by the layout of (ignored) objects \cite{moore1997perception, huang2007boolean, yu2019gestalt, yu2019similarity, wood2019processing, trick1997clusters}, which is similar to the kinds of patterns that visualizations are meant to reveal.

Demonstrating that inattentional blindness extends to data visualizations would underscore that understanding visualized data can require dynamic exploration, as opposed to instantaneous processing in a single glance. In a communication context, this would support the prescription that visualizations should be designed to guide viewers toward the ``right" pattern in a dataset. In an exploratory analysis context, it would suggest that important (or even surprising) patterns might be easily missed if they are not within the viewer’s current filter. Here, we present results from two experiments (a filtration task, and a corresponding control condition that omits the filter) that reveal robust inattentional blindness for a salient shape introduced to a set of ignored marks in a data visualization.

\begin{figure*}[h]
 \centering % avoid the use of \begin{center}...\end{center} and use \centering instead (more compact)
 \includegraphics[width=\textwidth]{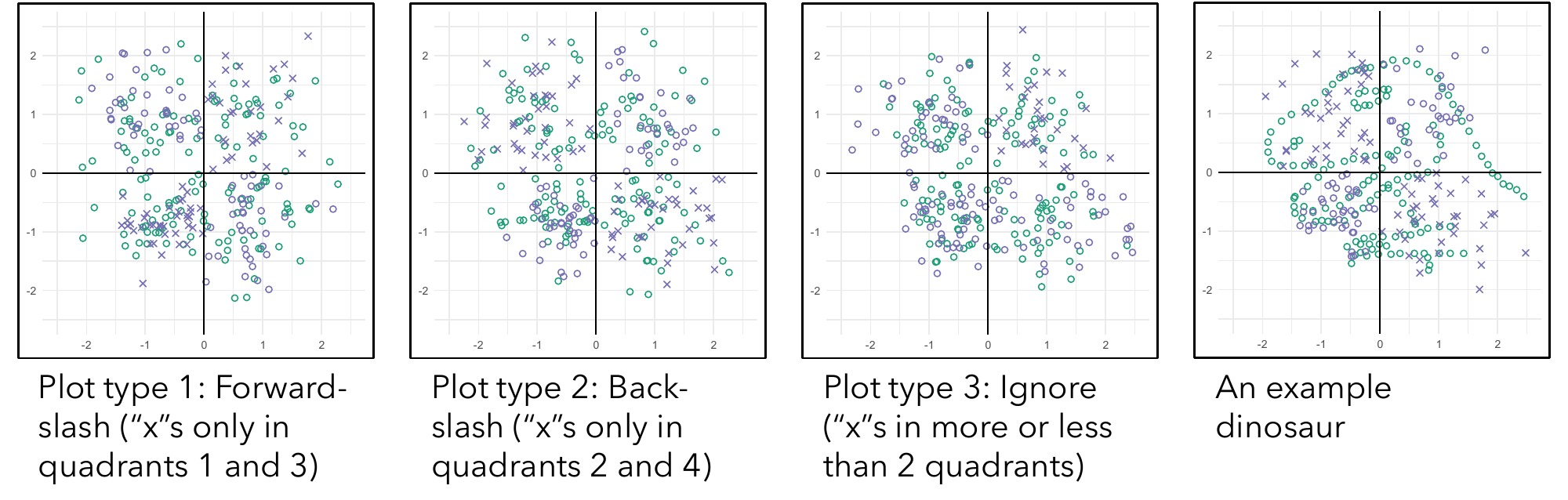}
 \caption{Types of plots shown in our tasks, along with an example dinosaur.}
 \label{fig:plots-fig}
\end{figure*}

Participants looked at a series of scatterplots in quick succession and completed a task that either required filtration (Experiment 1), or required simply looking at the scatterplots ``for anything unusual” (Experiment 2). Participants were not told that one set of points in the scatterplots occasionally formed a conspicuous dinosaur shape (as seen in \autoref{fig:plots-fig}). After completing the task, participants were asked whether they noticed anything unusual, such as patterns forming in certain groups, and then specifically asked whether they saw a dinosaur.

When faced with the filtration task, participants missed the dinosaur at an alarming rate. However, in the control task that omitted the filtration requirement, participants consistently saw the dinosaur, meaning that inattentional blindness extends to a data visualization in which a particular configuration of points is otherwise plainly clear. These filtering operations are surprisingly powerful, suggesting that, to ensure a more complete understanding of the patterns within a visualized dataset, visualizations must be explored over time, instead of bring processed instantly. 

\section{Related work}

Now a staple of the research literature on attention and perception, inattentional blindness was greeted with some astonishment even from seasoned experts in the field when it began to be widely reported in the 1990s \cite{mack1998inattentional, simons1999gorillas}. Earlier versions of the phenomenon, referred to as ``selective looking", date back to the 1970s; \cite{becklen1983selective, neisser1975selective, neisser2019control}.) In a classic set of studies, people reported whether the horizontal or vertical portion of a cross was longer on each trial and failed to notice even highly conspicuous - but unexpected - items that appeared in one of the cross’s quadrants on a critical trial \cite{mack1998inattentional}. The astonishment over such findings stemmed not only from the fact that they seemed to run counter to lay intuition that seeing is a function merely of the eyes, but also from the fact that much evidence until then had suggested that some salient features should be noticed without the need for attention \cite{treisman1988feature}. 

Work on inattentional blindness has shown both of these assumptions to be incorrect. For example, using eye-trackers, researchers found that people often fail to see the unexpected objects despite looking directly at them \cite{beanland2010looking, koivisto2004effects, memmert2006effects}. Building on a particularly well-known demonstration of inattentional blindness, where participants failed to see a person in a full-body gorilla outfit walk through the middle of a videotaped group of people passing basketballs \cite{simons1999gorillas}, one study found that people who missed the gorilla looked at it for about the same amount of time as did the people who saw the gorilla \cite{memmert2006effects}. 

Meanwhile, notions that some visual features should be seen without attention had grown out of research on ``visual search", in which participants sought targets that were embedded in arrays containing varying numbers (or ``set sizes") of non-target elements. Although it typically takes participants longer to find the target when it is embedded within a large array than within a small array, targets that were defined by a particularly unique or salient feature tended to be found quickly regardless of the size of the array, suggesting that they bypassed the need for attention, a phenomenon known as ``visual pop-out" \cite{treisman1988feature, wolfe1994guided}. Thus, subsequent findings that 28\% of participants were inattentionally blind to a red shape that was fully visible for 5 seconds among black and white shapes (one set of which participants were tracking; \cite{most2001not}) and that 32\% were inattentionally blind to a white shape visible for 5 seconds among uniformly black shapes \cite{most2005you} contrasted with assumptions that such unique and salient features should push their way into participants’ awareness \cite{mack1998inattentional}.

Why would features that appear to bypass the need for attention in visual search tasks fail to be noticed under conditions of inattentional blindness? A key difference between these tasks is that people know to expect the critical item in visual search tasks, whereas the critical item is unexpected in experiments on inattentional blindness. Thus, in visual search tasks, attention is broadly directed to the visual display in search for the target. Indeed, work on inattentional blindness has repeatedly underscored the importance of how people tune or direct their attention: when the unexpected item contains features similar to those that people are actively attending to, people notice it much more than when its features are similar to the items that people try to filter out. For example, when people tracked moving black items and ignored moving white items, they were far more likely to notice an unexpected and unique black than white shape that traversed the display, with the reverse pattern emerging among those who tracked the white items; in both cases, noticing of a gray shape fell in between \cite{most2001not, most2005you, simons1999gorillas}. In short, much of what we see depends on how our attention filters the world around us.

Research on visual attention has clear connections to visualization perception, which often requires attending to some trends and ignoring others. For example, change blindness -- a failure to detect salient changes to a scene due to capacity limits and comparison mechanisms -- presents powerful constraints for displays where viewers make visual comparisons between separate charts \cite{gleicher2011visual, nowell2001change}.

\section{Experiment 1: filtration task}

In Experiment 1, participants completed a task that required them to focus on a signal class of marks (blue X’s) in a scatterplot, while selectively filtering out blue and green circles. Based on existing demonstrations in the perceptual psychology literature, we hypothesized that this filtering requirement would cause sufficient inhibition of the other marks that participants would fail to notice the dinosaur shape.

\subsection{Stimuli}

All stimuli were created in R using ggplot2 \cite{ggplot2}. The scatterplots contained three classes of marks, as seen in \autoref{fig:plots-fig}. The dinosaurs were plotted as green circles. To maximize the incentive for viewers to selectively inhibit both the green color and circle shapes, we asked them to attend only to blue X's (to encourage inhibition of green and circles), and added a set of to-be-ignored blue circles (to additionally encourage inhibition of circle shapes).

We adapted the coordinates of the dinosaur shape from an image designed by Alberto Cairo \cite{cairo1970}, and centered their coordinates to have mean 0 and standard deviation 1 within our arbitrary coordinate space. Because people can detect the statistical properties of collections of objects even when those collections are not selectively attended (e.g., mean position; see \cite{alvarez2008representation}), we maximized the statistical similarity among the three sets of marks on the display. So, each group — including the green circles in plots where there was no dinosaur — was drawn from a normal distribution with the same mean, standard deviation, and number of samples in each quadrant as the dinosaur had.

\subsection{Task and procedure}

On each trial in this experiment, participants saw a series of seven graphs in quick succession. In Experiment 1a, each plot appeared on screen for 1s, and then disappeared for 250ms before the next one appeared. In Experiment 1b, we extended this time, such that plots stayed on screen for 2.5s, and then disappeared for 250ms.

In each series of seven plots, participants were asked to determine whether there were more plots in which the blue X's created a forward-slash (i.e., the blue X's were only in quadrants 1 and 3; plot type 1 in \autoref{fig:plots-fig}) or more plots in which the blue X's created a backward-slash (i.e., the blue X's were only in quadrants 2 and 4; plot type 2 in \autoref{fig:plots-fig}). To ensure participants could not ``cheat" by determining whether a plot created a forward-slash or back-slash by looking only at only the top half (or any other half) of the graphs, we created a third plot type, in which the blue X's were either in only one quadrant or in three quadrants (plot type 3 in \autoref{fig:plots-fig}).

Each series of seven plots contained either three plots with forward-slashes and two plots with back-slashes, or three plots with back-slashes and two plots with forward-slashes (chosen randomly for each participant). The remaining two plots were distractors (plot type 3); in one of the two, the blue X's were present only in one quadrant, and in the second the blue X's were present in three quadrants. The order of the plots in each sequence was randomized for each trial.

Participants completed five trials in which they saw the sequence of plots described above. On both the third and fifth trials, one of the plots in the sequence contained the dinosaur shape within the green circle marks. The dinosaur was randomly chosen to appear in any plot type (a forward-slash plot, back-slash plot, or neither). The position of the dinosaur within the third and fifth trial sequence was randomly chosen between the fourth and sixth plot of each of the sequences.

After completing the experiment, participants were asked whether they noticed any patterns in any of the groups of data points. Then, they were asked explicitly whether they noticed a dinosaur in any of the plots, producing our dependent variable. Finally, they were also asked whether they noticed a gorilla in any of the plots. However, no plots contained a gorilla — we added this question to account for the potential that participants would claim to notice things without actually seeing them.

Readers may try all tasks presented in this paper for themselves at \url{https://datasaurus-vis.herokuapp.com/}

\subsection{Participants}

For each of Experiment 1a and Experiment 1b, we recruited 20 participants from the online recruiting platform Prolific. For a discussion of the reliability of this subject pool, see \cite{peer2017beyond}. Participants were compensated upon completion of the experiment at a rate of \$9/hour.

\subsection{Results}

We excluded 1 participant in Experiment 1a who did not submit a complete dataset. We also excluded 4 participants in Experiment 1a, and 2 participants in Experiment 1b, whose accuracy in the given task was below chance (50\%), under the assumption that these participants were not paying attention to the task, or simply could not perform it.

These exclusion criteria left us with 15 participants in Experiment 1a, and 18 participants in Experiment 1b. In Experiment 1a, when the plots were visible for 1000ms, 14/15 (93.3\%) of participants missed the dinosaur. Furthermore, the single participant who claimed to see the dinosaur \textit{also} claimed to see the gorilla as well as other patterns, none of which were actually present. In Experiment 1b, when the plots were visible for a full 2.5s, 11/18 (61.1\%) of participants missed the dinosaur. Two of the participants who claimed to see the dinosaur also claimed to see other things (e.g. the gorilla) that were never present.

In both experiments, participants missed the dinosaur at high rates, suggesting that the filtering requirements of the scatterplot judgment task led to inattentional blindness in the data visualization.

\section{Experiment 2: control}

The results of Experiments 1a and 1b suggest that the filtering requirements of the task caused participants to miss the dinosaur. Yet, it is also possible that the dinosaur was simply too hard to see, perhaps overcrowded by other points in the display, or appeared too quickly to be perceived. Intuitively, viewing a trial suggests that neither is true, as the dinosaur is easily visible, and is present for a full 2.5s in Experiment 1b. Therefore, Experiments 2a and 2b test a control condition that relies on identical displays and timing, but selectively removes the filtering requirement of the task. If the filtering \textit{per se} were responsible for the missed dinosaur, then miss rates should fall substantially in the control conditions.

\subsection{Stimuli}

We used the same stimuli as in Experiments 1a and 1b.

\subsection{Task and procedure}

The plot sequence, randomization, and parameters here were exactly the same as they were in Experiments 1a and 1b. Participants again saw five trials in which a sequence of seven plots appeared, with the same randomization and plot types as before. In Experiment 2a, the plots appeared for 1000ms (as in Experiment 1a), and in Experiment 2b, they appeared for 2500ms (as in Experiment 1b). The dinosaur again appeared on the third and fifth trial sequences, with the same randomization as in 1a and 1b. The key difference was that participants were not told to count forward-slashes and back-slashes within the blue X's. Instead, they were told simply to watch the plots for ``anything unusual." As before, at the end of the experiment, participants were asked whether they noticed any trends, and whether they saw a dinosaur or gorilla.

\subsection{Participants}

For each of Experiment 2a and Experiment 2b, we recruited 20 new participants from Prolific.

\subsection{Results}

Because there was no participant task that could produce objectively correct answers, we did not exclude participants in these experiments based on accuracy, as we did before. We excluded 1 participant in Experiment 2b who did not submit a full dataset due to a technical error.

In Experiment 2a, when the plots were visible for 1s, 5/20 (25\%) of participants missed the dinosaur (one participant who claimed to see the dinosaur also claimed to see the gorilla), compared to Experiment 1a where 93.3\% of participants missed the dinosaur. In Experiment 2b, when the plots were visible for 2.5s, only 2/19 (10.5\%) of participants missed the dinosaur (five participants who claimed to see the dinosaur and also claimed to see the gorilla), compared to 61.1\% in Experiment 1b. There was a significant relationship between the rate of missing the dinosaur and the task type (i.e., filtration vs. control) both when comparing Experiment 1a to Experiment 2a ($\chi^2(1, N = 35) = 13.49, p<0.001$) and when comparing Experiment 1b to Experiment 2b ($\chi^2(1, N = 37) = 8.28, p<0.01$).

\begin{figure}[h]
 \centering % avoid the use of \begin{center}...\end{center} and use \centering instead (more compact)
 \includegraphics[width=0.6\columnwidth]{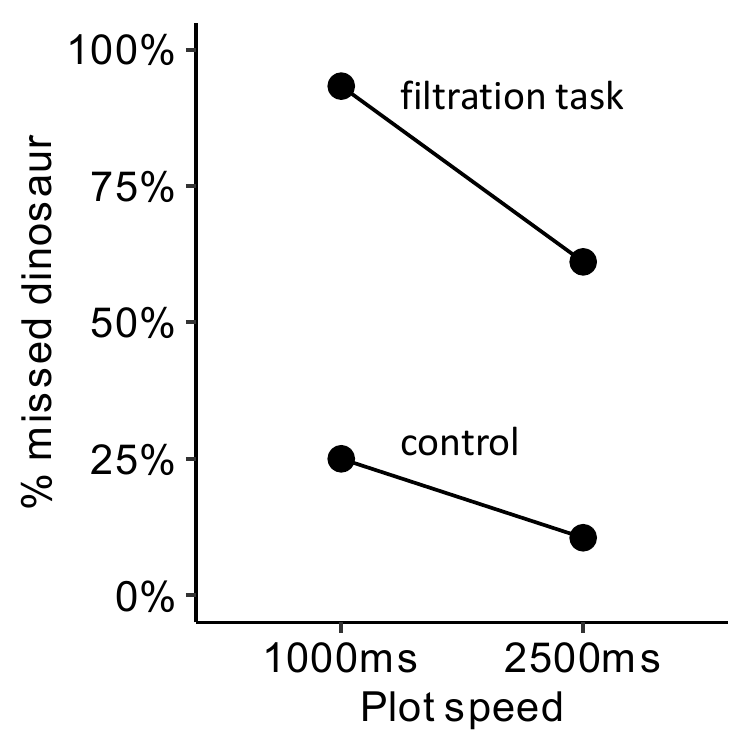}
 \caption{Rates of inattentional blindness (percent of participants who missed the dinosaur) by task and plot speed.}
 \label{fig:results-fig}
\end{figure}

\begin{figure*}[h]
 \centering % avoid the use of \begin{center}...\end{center} and use \centering instead (more compact)
 \includegraphics[width=\textwidth]{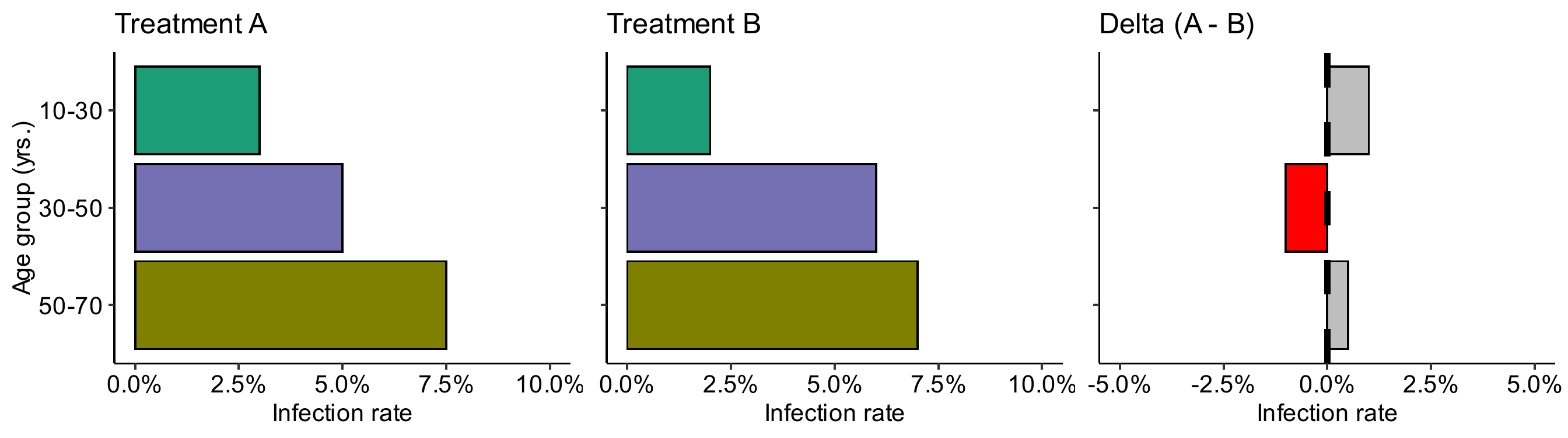}
 \caption{A plot showing how designers can effectively communicate data to limit filtering operations. Some designers might default to creating only the two left-most plots. However, including the third plot -- which shows differences between groups -- reduces the number of comparisons a reader needs to make and points them to the ``right" pattern in the data.}
 \label{fig:good-example-fig}
\end{figure*}

\section{Discussion \& future work}

In two experiments, we found that when inspecting a pattern in a single class of marks (blue X’s), the resulting visual filtering operation was so powerful that participants missed an otherwise clearly visible dinosaur shape made of relatively inhibited marks (green circles). %These results contribute to a growing psychology literature on how people interpret visual information over time \cite{franconeri2013nature, kawabata1986attention} while also presenting directions for future research.

\subsection{Implications for designing data visualizations}

%When participants were asked to make a judgment on a class of marks, their resulting filtering operation was so powerful that they failed to notice a dinosaur shape within one of the ignored classes of marks. 
The failure to perceive a salient and surprising pattern among a relatively ignored set of marks has practical implications for designing effective and informative data visualizations. First, visualization designers cannot assume that the same patterns that are salient to them will be salient for other viewers. In fact, recent work shows that viewers sometimes assume that others will focus on the same patterns that they do in a visualization, while in reality, each person can have their own idiosyncratic set of filters for what patterns they focus on \cite{xiong2019curse}. Designers might consider simplifying the dimensionality of the data that they show, reducing the space of possible filtering operations (and resulting comparisons) that a viewer needs to make, or at minimum, using annotation and highlighting to help a viewer see critical patterns instead of assuming that they will be noticed \cite{ajani2021declutter}.

For example, consider a designer who wants to show how two  treatments affect infection rates differently across age groups. Specifically, the designer wants to show that treatment A seems more effective than B in preventing infection for younger and older patients, but that B is more effective for patients 30-50 years old. One could show the data as two juxtaposed bar graph series, using only the left and center columns of \autoref{fig:good-example-fig}. However, that design requires the reader to make second-order comparisons of the (A-B) delta values for each age group. This requires multiple filtering operations that might be obvious to the visualization's designer, but not for the novice viewer \cite{xiongVisual}. A better design might facilitate extraction of the critical pattern by including a third series that depicts the deltas directly, as in the right-most plot of \autoref{fig:good-example-fig} \cite{nothelfer2019measures}.

\subsection{Directions for future work}

In the experiments presented here, the trend that participants missed was purely visual (the dinosaur formed by the green points). This presents a case where the trend missed is visually obvious and unique at a perceptual level, and should be remembered if it had been noticed. Another strategy for future work might be to seek similar cases where the surprising pattern is not perceptual, but conceptual, by showing that viewers can miss a relatively simple pattern that otherwise clearly invalidates an argument, or the validity of the dataset, in a way that should be easy to spot if the pattern were noticed (e.g., data revealing that a child appears to be older than their parent). Such a demonstration would help demonstrate the power of progressive visual filtering operations with a dependent measure more tied to the types of sensemaking and decisions that visualizations are meant to produce \cite{lee2015people}.

Future work may also explore inattentional blindness in data visualizations parametrically in ways beyond those presented here. Between our experiments, the only parameter that changed was the amount of time the plots stayed on the screen (1000ms in Experiments 1a and 2a, compared to 2500ms in Experiments 1b and 2b). However, one could also manipulate the number of distractor points, the number of total groups (i.e., adding or removing a distractor group entirely), or various visual features of the existing points (e.g., making the points for the dinosaur group thicker). Exploring this parameter space might shed light on the power and limits of visual filtering operations in realistic data displays.

\section{Conclusion}

Our experiments revealed consistent inattentional blindness for data visualization, lending strong support to the idea that data visualizations are explored over time rather than instantly seen. When tasked with performing a series of filtration operations over time, participants missed an obvious dinosaur at a significantly higher rate than when they were told to simply look at the plots for anything unusual. These results strengthen the evidence for the design guideline that visualization viewers should be guided toward the ``right" pattern in a communication context. Otherwise, if they fail to filter for the right type of information, even highly salient and important patterns can be missed.

\bibliographystyle{abbrv-doi}

\bibliography{ms}

\newcommand{\noop}[1]{}
\begin{thebibliography}{10}

\bibitem{ajani2021declutter}
K.~Ajani, E.~Lee, C.~Xiong, C.~N. Knaflic, W.~Kemper, and S.~Franconeri.
\newblock Declutter and focus: Empirically evaluating design guidelines for
  effective data communication.
\newblock {\em IEEE Transactions on Visualization and Computer Graphics}, 2021.

\bibitem{alvarez2008representation}
G.~A. Alvarez and A.~Oliva.
\newblock The representation of simple ensemble visual features outside the
  focus of attention.
\newblock {\em Psychological science}, 19(4):392--398, 2008.

\bibitem{beanland2010looking}
V.~Beanland and K.~Pammer.
\newblock Looking without seeing or seeing without looking? eye movements in
  sustained inattentional blindness.
\newblock {\em Vision research}, 50(10):977--988, 2010.

\bibitem{becklen1983selective}
R.~Becklen and D.~Cervone.
\newblock Selective looking and the noticing of unexpected events.
\newblock {\em Memory \& cognition}, 11(6):601--608, 1983.

\bibitem{brooks2015traditional}
J.~L. Brooks.
\newblock Traditional and new principles of perceptual grouping.
\newblock 2015.

\bibitem{cairo1970}
A.~Cairo.
\newblock Download the datasaurus: Never trust summary statistics alone; always
  visualize your data, Aug 2016.

\bibitem{franconeriCurr}
S.~Franconeri.
\newblock Three perceptual tools for seeing and understanding visualized data.
\newblock {\em Current Directions in Psychological Science}.
\newblock (in press).

\bibitem{franconeri2013nature}
S.~Franconeri.
\newblock The nature and status of visual resources.
\newblock In {\em The Oxford Handbook of Cognitive Psychology}. Oxford
  University Press, 2013.

\bibitem{franconeripspi}
S.~Franconeri, L.~Padilla, P.~Shah, J.~Zacks, and J.~Hullman.
\newblock The science of visual data communication: What works.
\newblock {\em Psychological Science in the Public Interest}.
\newblock (in press).

\bibitem{gleicher2011visual}
M.~Gleicher, D.~Albers, R.~Walker, I.~Jusufi, C.~D. Hansen, and J.~C. Roberts.
\newblock Visual comparison for information visualization.
\newblock {\em Information Visualization}, 10(4):289--309, 2011.

\bibitem{healey2011attention}
C.~Healey and J.~Enns.
\newblock Attention and visual memory in visualization and computer graphics.
\newblock {\em IEEE transactions on visualization and computer graphics},
  18(7):1170--1188, 2011.

\bibitem{huang2007boolean}
L.~Huang and H.~Pashler.
\newblock A boolean map theory of visual attention.
\newblock {\em Psychological review}, 114(3):599, 2007.

\bibitem{koivisto2004effects}
M.~Koivisto, J.~Hy{\"o}n{\"a}, and A.~Revonsuo.
\newblock The effects of eye movements, spatial attention, and stimulus
  features on inattentional blindness.
\newblock {\em Vision research}, 44(27):3211--3221, 2004.

\bibitem{lee2015people}
S.~Lee, S.-H. Kim, Y.-H. Hung, H.~Lam, Y.-a. Kang, and J.~S. Yi.
\newblock How do people make sense of unfamiliar visualizations?: A grounded
  model of novice's information visualization sensemaking.
\newblock {\em IEEE transactions on visualization and computer graphics},
  22(1):499--508, 2015.

\bibitem{mack1998inattentional}
A.~Mack and I.~Rock.
\newblock Inattentional blindness: Perception without attention.
\newblock In R.~D. Wright, ed., {\em Visual attention}, pp. 55--76. Oxford
  University Press, 1998.

\bibitem{memmert2006effects}
D.~Memmert.
\newblock The effects of eye movements, age, and expertise on inattentional
  blindness.
\newblock {\em Consciousness and cognition}, 15(3):620--627, 2006.

\bibitem{moore1997perception}
C.~M. Moore and H.~Egeth.
\newblock Perception without attention: Evidence of grouping under conditions
  of inattention.
\newblock {\em Journal of Experimental Psychology: Human Perception and
  Performance}, 23(2):339, 1997.

\bibitem{most2005you}
S.~B. Most, B.~J. Scholl, E.~R. Clifford, and D.~J. Simons.
\newblock What you see is what you set: sustained inattentional blindness and
  the capture of awareness.
\newblock {\em Psychological review}, 112(1):217, 2005.

\bibitem{most2001not}
S.~B. Most, D.~J. Simons, B.~J. Scholl, R.~Jimenez, E.~Clifford, and C.~F.
  Chabris.
\newblock How not to be seen: The contribution of similarity and selective
  ignoring to sustained inattentional blindness.
\newblock {\em Psychological science}, 12(1):9--17, 2001.

\bibitem{neisser2019control}
U.~Neisser.
\newblock The control of information pickup in selective looking.
\newblock In {\em Perception and its development}, pp. 201--219. Psychology
  Press, 2019.

\bibitem{neisser1975selective}
U.~Neisser and R.~Becklen.
\newblock Selective looking: Attending to visually specified events.
\newblock {\em Cognitive psychology}, 7(4):480--494, 1975.

\bibitem{nothelfer2019measures}
C.~Nothelfer and S.~Franconeri.
\newblock Measures of the benefit of direct encoding of data deltas for data
  pair relation perception.
\newblock {\em IEEE transactions on visualization and computer graphics},
  26(1):311--320, 2019.

\bibitem{nowell2001change}
L.~Nowell, E.~Hetzler, and T.~Tanasse.
\newblock Change blindness in information visualization: A case study.
\newblock In {\em Information Visualization, IEEE Symposium on}, pp. 15--15.
  IEEE Computer Society, 2001.

\bibitem{peer2017beyond}
E.~Peer, L.~Brandimarte, S.~Samat, and A.~Acquisti.
\newblock Beyond the turk: Alternative platforms for crowdsourcing behavioral
  research.
\newblock {\em Journal of Experimental Social Psychology}, 70:153--163, 2017.

\bibitem{shah2011bar}
P.~Shah and E.~G. Freedman.
\newblock Bar and line graph comprehension: An interaction of top-down and
  bottom-up processes.
\newblock {\em Topics in cognitive science}, 3(3):560--578, 2011.

\bibitem{simons1999gorillas}
D.~J. Simons and C.~F. Chabris.
\newblock Gorillas in our midst: Sustained inattentional blindness for dynamic
  events.
\newblock {\em perception}, 28(9):1059--1074, 1999.

\bibitem{szafir2016four}
D.~A. Szafir, S.~Haroz, M.~Gleicher, and S.~Franconeri.
\newblock Four types of ensemble coding in data visualizations.
\newblock {\em Journal of vision}, 16(5):11--11, 2016.

\bibitem{treisman1988feature}
A.~Treisman and S.~Gormican.
\newblock Feature analysis in early vision: evidence from search asymmetries.
\newblock {\em Psychological review}, 95(1):15, 1988.

\bibitem{trick1997clusters}
L.~M. Trick and J.~T. Enns.
\newblock Clusters precede shapes in perceptual organization.
\newblock {\em Psychological Science}, 8(2):124--129, 1997.

\bibitem{ggplot2}
H.~Wickham.
\newblock {\em ggplot2: Elegant Graphics for Data Analysis}.
\newblock Springer-Verlag New York, 2016.

\bibitem{wolfe1994guided}
J.~M. Wolfe.
\newblock Guided search 2.0 a revised model of visual search.
\newblock {\em Psychonomic bulletin \& review}, 1(2):202--238, 1994.

\bibitem{wood2019processing}
K.~Wood and D.~J. Simons.
\newblock Processing without noticing in inattentional blindness: A replication
  of moore and egeth (1997) and mack and rock (1998).
\newblock {\em Attention, Perception, \& Psychophysics}, 81(1):1--11, 2019.

\bibitem{xiongVisual}
C.~Xiong, V.~Setlur, B.~Bach, K.~Lin, E.~Koh, and S.~Franconeri.
\newblock Visual arrangements of bar charts influence comparisons in viewer
  takeaways.
\newblock {\em IEEE VIS 2021}.
\newblock (in press).

\bibitem{xiong2019curse}
C.~Xiong, L.~Van~Weelden, and S.~Franconeri.
\newblock The curse of knowledge in visual data communication.
\newblock {\em IEEE transactions on visualization and computer graphics},
  26(10):3051--3062, 2019.

\bibitem{yu2019gestalt}
D.~Yu, D.~Tam, and S.~L. Franconeri.
\newblock Gestalt similarity groupings are not constructed in parallel.
\newblock {\em Cognition}, 182:8--13, 2019.

\bibitem{yu2019similarity}
D.~Yu, X.~Xiao, D.~K. Bemis, and S.~L. Franconeri.
\newblock Similarity grouping as feature-based selection.
\newblock {\em Psychological Science}, 30(3):376--385, 2019.

\bibitem{zacks2020designing}
J.~M. Zacks and S.~L. Franconeri.
\newblock Designing graphs for decision-makers.
\newblock {\em Policy Insights from the Behavioral and Brain Sciences},
  7(1):52--63, 2020.

\end{thebibliography}
\end{document}